\documentclass[prd,showkeys,floatfix,nofootinbib,
               preprint,12pt,
               tightenlines,fleqn]{revtex4} 

\usepackage{amsmath,amssymb,revsymb,graphicx,dcolumn}
\usepackage{leftidx}

\newcommand{\beq}{\begin{equation}}
\newcommand{\eeq}{\end{equation}}
\newcommand{\beqa}{\begin{eqnarray}}
\newcommand{\eeqa}{\end{eqnarray}}
\newcommand{\bsubeqs}{\begin{subequations}}
\newcommand{\esubeqs}{\end{subequations}}

\usepackage{enumerate}

\begin{document}
\title[]
      {A note on bulk locality and covariance in AdS/CFT\vspace*{5mm}}
\author{Slava Emelyanov}
\email{viacheslav.emelyanov@physik.uni-muenchen.de}
\affiliation{Arnold Sommerfeld Center for
Theoretical Physics,\\
Ludwig Maximilian University (LMU),\\
80333 Munich, Germany\\}

\begin{abstract}
\vspace*{2.5mm}\noindent
This paper studies a recently proposed relation between the emergence of bulk locality in AdS/CFT
and the theory of quantum error correction. We show that if this relation is indeed realized in AdS/CFT, 
then bulk covariance is broken in the semi-classical limit.

\end{abstract}


\keywords{locality, covariance, effective quantum field theory, AdS/CFT correspondence}
\date{\today}

\maketitle

\section{Introduction}

According to the ``extrapolate" dictionary of the anti-de Sitter/conformal field theory (AdS/CFT)
correspondence, one can construct unbounded bulk CFT operators from the boundary ones, i.e. 
\beqa\label{eq:dictionary}
\hat{\phi}(x) &=& \int dX\,K(x,X)\,\hat{\mathcal{O}}(X)\,,
\eeqa
where $K(x,X)$ is the so-called ``smearing function", $x \in \mathcal{M}$ and $X \in \partial\mathcal{M}$~\cite{Hamilton&Kabat&Lifschytz&Lowe}.
One can further define an algebra $\mathcal{O}(\mathcal{M})$ generated by polynomials of the
bulk CFT operators smeared over test functions $\{f(x)\}$, i.e. functions which are smooth and compactly supported. 
Thus, one may consider the bulk CFT operator algebra $\mathcal{O}(\mathcal{M})$
instead of the boundary CFT algebra $\mathcal{O}(\partial\mathcal{M})$ inside of AdS space 
$\mathcal{M}$.\footnote{Note that the bulk operators may also be treated as elements of an enlarged CFT
algebra composed of $\mathcal{O}(\partial\mathcal{M})$ and $\mathcal{O}(\mathcal{M})$. This is not of a fundamental importance for our discussion below.}

On the other hand, one may consider a semi-classical quantum field theory in AdS space $\mathcal{M}$. 
For simplicity,
we choose a linear scalar field model characterized by an unbounded field operator $\hat{\Phi}(x)$, where 
$x \in \mathcal{M}$. One can define another algebra, i.e. $\mathcal{A}(\mathcal{M})$, generated by polynomials 
of smeared (over $\{f(x)\}$) scalar field 
operators. This algebra is supposed to satisfy the basic principles (axioms) of quantum field theory,
among of which are {\sl locality}, i.e. $[\hat{\Phi}(x),\hat{\Phi}(y)] = 0$,
whenever $x$ and $y$ are space-like separated, and {\sl covariance} roughly meaning that the algebraic 
structure of $\mathcal{A}(\mathcal{M})$ does not change under diffeomorphisms (see, for instance,~\cite{Brunetti&Fredenhagen,Fewster&Verch}).

In the semi-classical limit of quantum gravity, the bulk CFT theory should reduce to an effective quantum field
theory on a classical geometric background. Hence, one may expect the bulk CFT operator algebra 
$\mathcal{O}(\mathcal{M})$ satisfies the standard axioms of local quantum field theory. However,
it has been recently argued that $\mathcal{O}(\mathcal{M})$ is not local even at the semi-classical
approximation~\cite{Almheiri&Dong&Harlow,Harlow}. The argument is based on the following assumptions:
\begin{enumerate}
\item the time-slice axiom holds for $\mathcal{O}(\mathcal{M})$;
\item $\mathcal{O}(\mathcal{M})$ is complete (or irreducible).
\end{enumerate}

This argument is further employed in~\cite{Almheiri&Dong&Harlow,Harlow} to suggest a relation between the quantum 
error correction theory and the reproduction of the bulk in AdS/CFT. However, if this is more than an analogy, then it is 
problematic to have covariance in the bulk. This is our motivation to revisit the argument of~\cite{Almheiri&Dong&Harlow,Harlow}. 

The outline of the paper is as follows. In Sec.~\ref{sec:bl}, we show that the algebra $\mathcal{O}(\mathcal{M})$
could be local in the semi-classical limit. We clarify then the meaning of the assumptions made in~\cite{Almheiri&Dong&Harlow,Harlow} 
and explain why the argument of~\cite{Almheiri&Dong&Harlow,Harlow} 
{\sl cannot} lead to bulk non-locality at the level of the operator algebra. In Sec.~\ref{sec:bc}, we discuss
bulk covariance and point out that if the quantum error correction theory is indeed encoded in AdS/CFT as 
envisaged by~\cite{Almheiri&Dong&Harlow,Harlow}, then bulk covariance must be broken. In Sec.~\ref{sec:cr}, 
we provide final concluding remarks.

\section{Bulk locality}
\label{sec:bl}

\subsection{Bulk locality of CFT operators}
\label{sec:a}

It has been recently argued~\cite{Almheiri&Dong&Harlow,Harlow} that bulk locality of the bulk CFT
operators is not respected at the level of the operator algebra (i.e. in the strong sense), but only at the level of
certain matrix elements (i.e. in the weak sense). The statement is made in the semi-classical limit. If it is correct, then a certain
$\mathcal{O}(\mathcal{M})$ theory does not reduce to a certain local quantum field theory $\mathcal{A}(\mathcal{M})$ in the bulk, 
which is supposed to be an effective field theory in the low-energy limit.

However, the bulk CFT algebra could be local in the \emph{strong} sense. Indeed, in the semi-classical limit, we have
\beqa\label{eq:commutator1}
[\hat{\phi}(x),\hat{\phi}(y)] &=& i\int dXdY\,K(x,X)\,K(y,Y)\,\Delta_{\mathcal{O}}(X,Y)\,,
\eeqa
where $[\hat{\mathcal{O}}(X),\hat{\mathcal{O}}(Y)] \equiv i\Delta_{\mathcal{O}}(X,Y)$. It is worth emphasizing that
$\Delta_{\mathcal{O}}(X,Y)$ does not depend on a quantum state, i.e. it is state-independent. For simplicity, we consider the
Poincar\'e patch of the three-dimensional AdS space, i.e. $\mathcal{M} = \text{PAdS}_3$. The commutator of the CFT operator at two boundary points is given by
\beqa
[\hat{\mathcal{O}}(T_x,X_x),\hat{\mathcal{O}}(T_y,X_y)] &\propto& 
\frac{1}{\left((\Delta T - i\varepsilon)^2 - \Delta X^2\right)^\Delta}
-\frac{1}{\left((\Delta T + i\varepsilon)^2 - \Delta X^2\right)^\Delta}\,,
\eeqa
where $\Delta T \equiv T_x - T_y$, $\Delta X \equiv X_x - X_y$, and $\Delta$ is the conformal weight of the CFT operator.
To further simplify computations, we set $y = (T_y,X_y,Z_y) = 0$ and $\Delta = 2$. Following~\cite{Hamilton&Kabat&Lifschytz&Lowe},
one can obtain
\beqa\label{eq:commutator2}
[\hat{\phi}(x),\hat{\phi}(0)] &=& \frac{Z_x^2}{2\pi}\left(\frac{1}{\left((T_x-i\varepsilon)^2 - X_x^2 - Z_x^2\right)^2}
-\frac{1}{\left((T_x+i\varepsilon)^2 - X_x^2 - Z_x^2\right)^2}\right).
\eeqa
Thus, the operator $\hat{\phi}(x)$ at $T_x = 0$ commutes with $\hat{\phi}(0)$ which is $\hat{\mathcal{O}}(0)$ up to the rescaling $1/Z_y^2$ in the limit $Z_y \rightarrow 0$. 

This result can be reproduced from $[\hat{\Phi}(x),\hat{\Phi}(0)]$ if the scalar field satisfies the massless Klein-Gordon equation with minimal coupling to gravity.
This is completely consistent with the conformal weight of the CFT operator and the dimension of the AdS geometry. Thus,
$[\hat{\phi}(x),\hat{\phi}(y)] $ does commute for $x$ and $y$ space-like separated in the \emph{strong} sense. This is also consistent with an expectation that
the low-energy limit of quantum gravity corresponds to the semi-classical quantum field theory.\footnote{See also the last par. of
sec. II.D in the third reference of~\cite{Hamilton&Kabat&Lifschytz&Lowe}}
Consequently, the argument of~\cite{Almheiri&Dong&Harlow,Harlow} based on a combination of the time-slice axiom and the completeness axiom should be revisited. 

\subsection{Time-slice and completeness axiom}

The time-slice axiom states that if $O$ is any fixed neighbourhood of a Cauchy surface 
$\Sigma$ of a globally hyperbolic space $\mathcal{M}$, then $\mathcal{A}(\mathcal{M})$ is generated by the scalar field operators 
having a non-vanishing support inside of $O$. In other words,  $\mathcal{A}(\mathcal{M}) \cong \mathcal{A}(O)$,
whenever $\Sigma \subset O \subset \mathcal{M}$~\cite{Haag&Schroer,Haag}.

The completeness axiom can be formulated in different ways~\cite{Haag&Schroer}. It is worth emphasizing that this axiom is related 
to a {\sl representation} of the algebra $\mathcal{A}(\mathcal{M})$ on a certain state $|\Omega\rangle$.\footnote{If one specifies a state
defined on $\mathcal{A}(\mathcal{M})$, then one can construct a Hilbert space representation of $\mathcal{A}(\mathcal{M})$
associated with this state. This is achieved through the so-called Gelfand-Naimark-Segal construction~\cite{Haag}.} 
In other words, this axiom imposes a certain constraint on a state. We will denote the algebra
representation as $\mathcal{A}_{\pi}(\mathcal{M})$ which is defined on a Hilbert space $\mathcal{H}$. The representation $\pi$ of
$\mathcal{A}(\mathcal{M})$, i.e.
$\mathcal{A}_{\pi}(\mathcal{M})$, is said to be irreducible if there is no a non-trivial bounded operator
commuting with all operators from $\mathcal{A}_{\pi}(\mathcal{M})$. If the representation is irreducible, then the set of operators 
$\mathcal{A}_{\pi}(\mathcal{M})$ is said to be complete.\footnote{Note that the authors of~\cite{Streater&Wightman} refer to~\cite{Ruelle} 
when they introduce a notion of irreducibility (completeness) of a set of operators, while the author of~\cite{Ruelle} seems to refer 
to~\cite{Haag&Schroer}.}

The completeness axiom and time-slice axiom allow to uniquely determine a quantum state by measurements 
performed in a small time interval. If the former is violated, then one cannot uncover all properties of the state.
If the latter is not fulfilled by a certain field theory, then one needs to observe the state at all times to determine
its properties~\cite{Haag&Schroer}.

It is worth mentioning that a more profound property of a representation of the operator algebra is its cyclicity 
(see Note added in proof in~\cite{Haag&Schroer}). A representation $\pi$ is cyclic if 
$\mathcal{A}_{\pi}(\mathcal{M})|\Omega\rangle$ is dense in $\mathcal{H}$. A state generating $\pi$ is then said to
be cyclic.

An example of the reducibility of the Hilbert space or incompleteness of the operator algebra should
clarify these. Suppose one chooses a factorized 
Hilbert space representation $\mathcal{H}_\text{L}{\otimes}\mathcal{H}_\text{R}$ of the total algebra 
$\mathcal{A}_\text{T}(M)$ in the eternal Schwarzschild black-hole geometry $M$. The ``right" operator algebra
$\mathcal{A}_\text{R}(M)$ (having a vanishing support in the ``left" outside region of the hole)
is incomplete in this representation, because the ``left" operator algebra $\mathcal{A}_\text{L}(M)$ (having a 
vanishing support in the ``right" outside region of the hole) is non-trivial. 
However, the physical vacuum, i.e. the Hartle-Hawking one,
generates an irreducible representation $\mathcal{H}_\text{T}$ of the algebra, because the Hartle-Hawking state
is cyclic with respect to it~\cite{Kay}. Specifically, $\mathcal{A}_\text{R}(M)$ (or $\mathcal{A}_\text{L}(M)$) acting on the
Hartle-Hawking state generates a space being {\sl dense} in the Hilbert space $\mathcal{H}_\text{T}$. It is worth emphasizing 
at this point that the algebras $\mathcal{A}_\text{L}(M)$ and $\mathcal{A}_\text{R}(M)$ commute independent on the representation
(cf.~\cite{Almheiri&Dong&Harlow,Harlow}).

Thus, we have two representations of local (causal) quantum field theory. One of the representations is reducible, while
another is irreducible. If one wants the completeness axiom to be fulfilled by the quantum theory, one needs to 
choose the irreducible representation. As it should be evident this has nothing to do with the locality axiom.

It is worth noting that these two representations from our example are not unitarily equivalent (for more
on this, see~\cite{Emelyanov}). We discuss analogous (reducible and irreducible) representations 
from a different perspective in the eternal AdS black-hole geometries in~\cite{Emelyanov1}.

It should be also evident that locality (causality) and the time-slice property of $\mathcal{A}(\mathcal{M})$ are state-independent, 
whereas completeness is state-dependent. These three axioms are a part of the postulates of local quantum field theory (see, for instance,~\cite{Haag}). They are not inconsistent with each other in the field model we consider in the bulk.\footnote{Note, these 
axioms are not satisfied {\sl a priori} in a certain quantum field theory. These have to be checked. We are guided by these principles in order to have a non-pathological quantum field theory.} As we have shown above, locality is also preserved at the level of the operator algebra 
$\mathcal{O}(\mathcal{M})$ in the semi-classical limit. 

Nevertheless, there could be a certain CFT theory on the boundary, which is not local 
in the bulk. However, such a theory would not reduce to any semi-classical quantum field theory in the 
low-energy limit of quantum gravity.

\subsection{Hilbert space representation and locality}

In Sec.~\ref{sec:a}, we have {\sl not} used any Hilbert space representation of the operators. In other words, the equations \eqref{eq:commutator1} and \eqref{eq:commutator2} are {\sl operator} equalities. To put it differently, 
this holds for any CFT state from the CFT Hilbert space. Note that if the backreaction in a certain CFT state is not negligible, then the whole geometry changes. This leads in particular to change of the smearing function as well 
as the commutator, because these depend on spacetime metric. In general, one cannot {\sl a priori} say anything about locality in the new geometry. However, this is certainly beyond of what is considered 
in~\cite{Almheiri&Dong&Harlow,Harlow}.\footnote{Note that it is \emph{not} excluded that there could be a certain
asymptotically AdS space $\bar{\mathcal{M}}$ such that the bulk algebra $\mathcal{O}(\bar{\mathcal{M}})$ (or the enlarged
algebra) is non-local in the \emph{strong} sense in the low-energy limit. However, as shown
above, this is \emph{not} the case in AdS geometry, wherein the author of~\cite{Almheiri&Dong&Harlow}
has however argued that the algebra is local \emph{only} in the \emph{weak} sense.} 

In order to discuss the completeness or irreducibility axiom, one should introduce a quantum state $|\omega\rangle$. 
We assume that this state is the ordinary CFT vacuum. This state generates a Hilbert space $\mathcal{H}_\omega$ through 
the Gelfand-Naimark-Segal procedure (see, for instance,~\cite{Haag}). This is the ordinary CFT Hilbert space.
A representation $\pi_\omega$ of $\mathcal{O}(\mathcal{M})$ is a set of linear operators acting on $\mathcal{H}_\omega$.
This representation of the operator algebra will be denoted as $\mathcal{O}_\pi(\mathcal{M})$ which is 
$\pi_\omega(\mathcal{O}(\mathcal{M}))$. It is worth emphasizing that the representation $\pi_\omega$ of the algebra 
does not change the algebraic structure of it -- $\pi_\omega$ is homomorphism. In other words, if $\mathcal{O}(\mathcal{M})$ is
local (causal), then $\mathcal{O}_\pi(\mathcal{M})$ is automatically local as well.

The representation $\pi_\omega$ could be of various types: faithful, cyclic, irreducible and so on~\cite{Haag}. We are mainly 
interested in understanding the irreducibility/reducibility property of $\pi_\omega$ in light of the argument of~\cite{Almheiri&Dong&Harlow,Harlow}. As noted above, the irreducibility can be defined in different ways~\cite{Haag&Schroer,Haag}. Following~\cite{Haag}, the irreducibility of $\pi_\omega$ means that no non-trivial element $\hat{o}$ of $\mathcal{O}(\mathcal{M})$ 
is represented on $\mathcal{H}_\omega$ by a trivial operator. In other words, if the representation $\pi_\omega$ is irreducible, 
then  $\pi_{\omega}(\hat{o}) \equiv \hat{o} _\omega \not\propto \hat{\mathbf{1}}$ for any $\hat{o} \not\propto \hat{\mathbf{1}}$ 
belonging to $\mathcal{O}(\mathcal{M})$. 

As pointed out above, this property of the representation is independent on the locality axiom and vice verse, because, for instance, 
locality is an operator 
statement. In other words, one {\sl cannot} use one of these axioms to prove or disprove another. The confusion apparently 
appears when one employs an {\sl alternative} formulation of what the irreducibility/completeness axiom means. Specifically,
if the representation $\pi_\omega$ is irreducible, then a bounded operator $\hat{B}$ commuting with all (unbounded) 
operators of $\mathcal{O}_\pi(\mathcal{M})$ is a multiple of the identity. Can one use this to prove non-locality of 
$\mathcal{O}(\mathcal{M})$? No, otherwise one is coming dangerously close to
proving the inconsistency of the Wightman axioms.
First, the operator $\hat{\phi}_\omega(x)$ is unbounded.\footnote{It is worth noting that the quantum field 
$\hat{\phi}(x)$ smeared out over a test function $f(x)$ is still unbounded operator, although the test function is bounded as this follows from its definition and the Weierstrass theorem.}
Second, when one asks whether
$[\hat{\phi}_\omega(x),\hat{\mathcal{O}}_\omega(X)]$ is vanishing or not for space-like separated $x$ and $X$, one
should bear in mind that $\hat{\phi}_\omega(x)$ is defined through \eqref{eq:dictionary}. Hence, one should instead ask
whether
\beqa
\int dY\,K(x,Y)\,[\hat{\mathcal{O}}_\omega(Y),\hat{\mathcal{O}}_\omega(X)] 
\eeqa
vanishes whenever $x$ and $X$ are space-like separated. The commutator of the CFT operators does not
depend on a state in the semi-classical limit. Therefore, one can omit the index $\omega$ which refers to the representation. The answer depends thus on the smearing function, rather than on the completeness/irreducibility axiom. As shown above, it does vanish at the level of the operator algebra, at least in the case we have considered 
in Sec.~\ref{sec:a}.

To sum it up, we disagree with the \emph{general} statement made in~\cite{Almheiri&Dong&Harlow} that bulk 
locality cannot be respected within the CFT \emph{at the level of the algebra of operators} (i.e. in the strong sense). Our argument is
based on two observations: First, one cannot prove non-locality of the theory employing the completeness 
and time-slice axioms. Second, the bulk operator algebra is local in the \emph{strong} sense in the semi-classical limit at least in AdS geometry.

\section{Bulk covariance}
\label{sec:bc}

\subsection{Covariance}

The three-dimensional AdS geometry is a hyperboloid embedded in space $\mathbf{R}_{2,2}^{4}$ with the line 
element $ds^2 = \eta_{ab}dx^adx^b$, where $a,b$ run from $0$ to $3$ and $\eta_{ab} = \text{diag}(+,-,-,+)$.

One may choose various coordinates which can cover the whole hyperboloid or merely a certain part of it.
We will consider the Poincar\'e patch $\mathcal{M}$ mentioned above and the AdS-Rindler patch, which
is denoted as $\mathcal{N}$, i.e. $\mathcal{N} = \text{RAdS}$, in the following. The RAdS patch is mapped to the Poincar\'{e}
one by a hyperbolic embedding $\psi$, namely $\psi\,{:}\,\, \mathcal{N} \rightarrow \mathcal{M}$.\footnote{A
hyperbolic embedding $\psi$ is an isometry, which preserves time and space orientation. An isometry is a diffeomorphism
such that $\psi_*g_{\mathcal{N}} = g_{\mathcal{M}}|_{\psi(\mathcal{N})}$, where, for instance, $g_{\mathcal{N}}$ is a metric
tensor in $\mathcal{N}$.} In other words, this map is specified via
expressing the Rindler-AdS coordinates through the Poincar\'e ones (see, for instance,~\cite{Griffiths&Podolsky}).

We have defined the algebra $\mathcal{A}(\mathcal{M})$ above. We denote the scalar field operator belonging to this 
algebra as $\hat{\Phi}_\mathcal{M}(x)$, where 
$x \in \mathcal{M}$ as before. One can also define an operator algebra
$\mathcal{A}(\mathcal{N})$, which is associated with the Rindler patch of AdS space. The field operator 
$\hat{\Phi}_{\mathcal{N}}(\tilde{x})$ refers to $\mathcal{A}(\mathcal{N})$, where $\tilde{x} \in \mathcal{N}$.
By covariance one understands
\beqa
\alpha_{\psi}\circ\hat{\Phi}_\mathcal{N} &=& \hat{\Phi}_\mathcal{M}\circ\psi_*\,,
\eeqa
where $\psi_*$ is a push-forward of $\psi$ which maps test functions $\{f(\tilde{x})\}$ to $\{(\psi_*f)(x)\}$, where $x = \psi(\tilde{x})$. 
The map $\alpha_{\psi}$ is an injective homomorphism\footnote{A homomorphism is, roughly speaking, a map which respects 
the algebraic structure.}
from $\mathcal{A}(\mathcal{N})$ to $\mathcal{A}(\mathcal{M})$~\cite{Brunetti&Fredenhagen,Fewster&Verch}.

To put it differently, we now consider points $p$, which lie in both $\mathcal{M}$ and $\mathcal{N}$, i.e. 
$p \in \mathcal{N}{\cap}\mathcal{M}$. For this set of points we have field operators $\hat{\Phi}_{\mathcal{M}}(x)$ and 
$\hat{\Phi}_{\mathcal{N}}(\tilde{x})$, such that
$p$ is parametrized by both $\{x\}$ and $\{\tilde{x}\}$. The covariance principle implies
\beqa
\hat{\Phi}_{\mathcal{N}}(\tilde{x}) &=& \hat{\Phi}_{\mathcal{M}}(\psi(\tilde{x}))\,.
\eeqa

\subsection{Bulk CFT operator reconstructions and covariance}

One may construct the bulk CFT operator for points $p$ either in $\mathcal{N}$ or 
$\mathcal{M}$~\cite{Hamilton&Kabat&Lifschytz&Lowe}. Thus, one can have two bulk CFT operators
$\hat{\phi}_\mathcal{M}(x)$ and $\hat{\phi}_\mathcal{N}(\tilde{x})$ at the same bulk points $p$. The bulk operator
$\hat{\phi}_\mathcal{M}(x)$ corresponds to the Poincar\'e or global reconstruction, while $\hat{\phi}_\mathcal{N}(\tilde{x})$ to the
AdS-Rindler reconstruction.

A counter-intuitive observation has been made in~\cite{Almheiri&Dong&Harlow,Harlow} based on 
different reconstructions
of the bulk CFT operators and the non-locality of $\mathcal{O}(\mathcal{M})$. Specifically, those two
bulk CFT reconstructions should {\sl not} be equivalent as operators, i.e. 
\beqa
\hat{\phi}_{\mathcal{N}}(\tilde{x}) &\neq& \hat{\phi}_{\mathcal{M}}(\psi(\tilde{x}))\,.
\eeqa
This observation was related to the gauge invariance of the boundary theory in~\cite{Mintun&Polchinski&Rosenhaus} 
(see also~\cite{Harlow}). However, the argument employed in~\cite{Almheiri&Dong&Harlow,Harlow} in favor of this inequivalence is generally invalid as we have shown above. Nevertheless, direct computations may show that they are indeed 
inequivalent.\footnote{It is worth noting that the Bogolyubov transformation is canonical, i.e. it does not change the 
commutator of the field operators. In other words, if $\hat{\phi}_{\mathcal{N}}(\tilde{x})$ and $\hat{\phi}_{\mathcal{M}}(x)$ are
related through the Bogolyubov transformation in $\mathcal{N}{\cap}\mathcal{M}$, then 
$\hat{\phi}_{\mathcal{N}}(\tilde{x}) = \hat{\phi}_{\mathcal{M}}(\psi(\tilde{x}))$ automatically.}

Suppose that $\hat{\phi}_\mathcal{N}(\tilde{x})$ and $\hat{\phi}_\mathcal{M}(x)$ are inequivalent as 
operators in $\mathcal{N}{\cap}\mathcal{M}$. This implies then that covariance is broken in the bulk. This is
unacceptable as it would be a pathological modification of effective quantum field theory. 

Moreover, string theory which is supposed to be dual to the CFT boundary theory is covariant. Its semi-classical limit provides still with a covariant effective field theory. Thus, the equation
\beqa
\hat{\phi}_{\mathcal{N}}(\tilde{x}) &=& \hat{\phi}_{\mathcal{M}}(\psi(\tilde{x}))\,
\eeqa
must hold at that limit.

\section{Concluding remarks}
\label{sec:cr}

The argument of~\cite{Almheiri&Dong&Harlow,Harlow} based on the time-slice and completeness axiom cannot judge 
whether a bulk CFT algebra composed of smeared bulk CFT operators (or the enlarged CFT algebra) taken at a certain time slice satisfies or does not
satisfy the locality (causality) axiom at the level of algebra of operators. We have explicitly shown an example of the bulk CFT operators which demonstrates 
this as well as an example when the completeness axiom is not fulfilled in a local (causal) theory.

The covariance principle is one of the basic axioms of general relativity which celebrates its 100th birthday 
this year. Thus, different reconstructions of bulk CFT operators should give the same operator at a given bulk 
point, at least in the low-energy limit. The contrary is employed in~\cite{Almheiri&Dong&Harlow} to make a contact of 
inequivalent bulk reconstructions with the quantum error correction theory. 
However, we would like to remind here that one should be careful when one applies the logic of quantum mechanics to quantum
field theory. If one does {\sl not} take into account differences between quantum mechanics and quantum field theory, then 
one can obtain the well-known paradoxical consequences. One of the paradoxes is posed in~\cite{Hegerfeldt}
and resolved in~\cite{Buchholz&Yngvason} (see also~\cite{Yngvason}).

\section*{
ACKNOWLEDGMENTS}

We are thankful to D. Harlow and D. Sarkar for discussions and their comments. 
It is also a pleasure to thank M. Haack and S. Konopka for discussions and their valuable suggestions/comments on an early version of this paper.
This research is supported by TRR 33 ``The Dark Universe''.


\end{document}